\documentclass[twocolumn]{aastex631}
\journalinfo{Published in The Astrophysical Journal Letters, 999:L16 (2026), DOI: 10.3847/2041-8213/ae4104}

\usepackage{graphicx,psfrag}
\usepackage{mathrsfs}
\usepackage{amsmath,amsfonts,amssymb}
\usepackage{multirow}
\usepackage{comment}
\usepackage[normalem]{ulem}
\usepackage{enumitem}
\usepackage{tcolorbox}
\usepackage{units}
\usepackage{xspace}
\usepackage[capitalise]{cleveref}
\usepackage[nohyperlinks]{acronym}
\usepackage{pifont}
\usepackage{soul}

\newcommand{\be}{\begin{equation}}
\newcommand{\ee}{\end{equation}}
\newcommand{\bea}{\begin{eqnarray}}
\newcommand{\eea}{\end{eqnarray}}
\newcommand{\bel}{\begin{align}}
\newcommand{\eel}{\end{align}}

\def\nic{{$^{56}$Ni}\xspace}
\def\cob{{$^{56}$Co}\xspace}

\def\thc{{\tt THC}\xspace}
\def\snec{{\tt SNEC}\xspace}
\def\skynet{{\tt SkyNet}\xspace}
\def\tardis{{\tt TARDIS}\xspace}

\def\DDa{{\tt DD2\_q1.77}\xspace}
\def\DDs{{\tt DD2\_q1.0}\xspace}
\def\SFH{{\tt SFHo\_q1.0}\xspace}
\def\BLh{{\tt BLh\_q1.49}\xspace}

\newcommand{\ie}{\textit{i.e.}\xspace}
\newcommand{\eg}{\textit{e.g.}\xspace}

\newacro{BNS}{binary neutron star}
\newacro{EOS}{equation of state}
\newacroplural{EOS}[EOSs]{equations of state}
\newacro{NS}{neutron star}
\newacro{BH}{black hole}
\newacro{KN}{kilonova}
\newacro{SN}{supernova}
\newacroplural{KN}[KNs]{kilonovae}
\newacroplural{SN}[SNe]{supernovae}
\newacro{NSE}{nuclear statistical equilibrium}
\newacro{NR}{numerical relativity}
\newacro{GW}{gravitational wave}

\definecolor{cyan}{rgb}{0,0.9,0.9}
\definecolor{orange}{rgb}{0.9,0.5,0}
\definecolor{magenta}{rgb}{1,0,1}
\definecolor{purple}{rgb}{0.8,0.4,0.8}
\definecolor{gray}{rgb}{0.8242,0.8242,0.8242}
\definecolor{light-gray}{gray}{0.95}

\soulregister\ac7
\soulregister\unit7
\soulregister\citet7
\soulregister\citep7
\soulregister\ie0
\soulregister\BLh0
\soulregister\SFH0
\soulregister\DDa0
\soulregister\DDs0
\soulregister\snec0
\soulregister\nic0
\soulregister\cob0

\makeatletter
\AtBeginDocument
 {
   \def\ltx@label#1{\cref@label{#1}}
   \def\label@in@display@noarg#1{\cref@old@label@in@display{#1}}
\def\label@in@mmeasure@noarg#1{%
    \begingroup%
      \measuring@false%
      \cref@old@label@in@display{#1}
    \endgroup}%
 } %
\makeatother

\begin{document}

\title{\nic production in long-lived binary neutron star merger remnants}

\author[0000-0001-8168-4579]{Maximilian Jacobi}
\email{maximilian.jacobi@uni-jena.de}
\affiliation{Theoretisch-Physikalisches Institut, Friedrich-Schiller-Universit{\"a}t Jena, 07743, Jena, Germany}
\author[0009-0005-0976-7851]{Fabio Magistrelli}
\affiliation{Theoretisch-Physikalisches Institut, Friedrich-Schiller-Universit{\"a}t Jena, 07743, Jena, Germany}
\author[0000-0001-7888-9733]{Eleonora Loffredo}
\affiliation{INAF - Osservatorio Astronomico d'Abruzzo, Via M.\ Maggini snc, 64100 Teramo, Italy}
\author[0000-0003-1626-1355]{Giacomo Ricigliano}
\affiliation{Max-Planck-Institut für Kernphysik, Saupfercheckweg 1, Heidelberg 69117, Germany}
\affiliation{Institut für Kernphysik, Technische Universität Darmstadt, Schlossgartenstr.\ 2, Darmstadt 64289, Germany}
\author[0000-0002-5112-1975]{Leonardo Chiesa}
\affiliation{Dipartimento di Fisica, Universit\'{a} di Trento, Via Sommarive 14, 38123 Trento, Italy}
\affiliation{INFN-TIFPA, Trento Institute for Fundamental Physics and Applications, via Sommarive 14, I-38123 Trento, Italy}
\author[0000-0002-2334-0935]{Sebastiano Bernuzzi}
\affiliation{Theoretisch-Physikalisches Institut, Friedrich-Schiller-Universit{\"a}t Jena, 07743, Jena, Germany}
\author[0000-0002-0936-8237]{Albino Perego}
\affiliation{Dipartimento di Fisica, Universit\'{a} di Trento, Via Sommarive 14, 38123 Trento, Italy}
\affiliation{INFN-TIFPA, Trento Institute for Fundamental Physics and Applications, via Sommarive 14, I-38123 Trento, Italy}
\author[0000-0002-6995-3032]{Almudena Arcones}
\affiliation{Institut für Kernphysik, Technische Universität Darmstadt, Schlossgartenstr.\ 2, Darmstadt 64289, Germany}
\affiliation{GSI Helmholtzzentrum für Schwerionenforschung GmbH, Planckstr.\ 1, Darmstadt 64291, Germany}
\affiliation{Max-Planck-Institut für Kernphysik, Saupfercheckweg 1, Heidelberg 69117, Germany}

\received{2025 November 4}
\revised{2026 January 28}
\accepted{2026 February 1}
\published{2026 February 26}

\begin{abstract}
  We investigate the nucleosynthesis and \acl{KN} emission based on numerical-relativity \acl{BNS} merger simulations that incorporate a two-moment neutrino-transport scheme.
  Unlike in previous works with simpler neutrino treatments, a massive, fast (up to $v=0.3c$), proton-rich neutrino-driven wind develops in the postmerger phase of the simulations as long as the merger remnant does not collapse to a \acl{BH}.
  We evolve the ejecta for 100 days after the merger using 2D ray-by-ray radiation-hydrodynamics simulations coupled in situ to a complete nuclear network.
  The most abundant nucleosynthesis products are He, \nic, and \cob.
  We find a total yield of $\sim 10^{-3} M_\odot$ of \nic for all mergers that produce massive \acl{NS} remnants, independently of the mass ratio and \aclp{EOS}.
  After a few days, the decay of \nic and later \cob becomes the primary source of heating in the matter expanding above the remnant.
  As a result, the \acl{KN} light curve flattens on timescales of days for polar observation angles.
  This is an important effect that should be included in future models of \acl{KN} light curves.
  Furthermore, the observation of a \nic--\cob feature could serve as a clear indicator for the presence of a long-lived \acl{NS} remnant in future \acl{KN} observations.
\end{abstract}

\section{Introduction}
The mergers of \ac{BNS} systems produce hot and dense ejecta, in which heavy elements are synthesized by the $r$-process \citep{Symbalisty:1982a,Eichler:1989ve,Freiburghaus:1999b}.
During the expansion of the ejecta, heat generated by the radioactive decay of the newly formed heavy elements powers a fast transient, the \ac{KN} \citep{Li:1998bw,Metzger:2010sy}.
The combined heating of many $r$-process nuclei decaying simultaneously typically leads to a power law--like heating rate $\propto t^{-\alpha}$ with $\alpha \approx 1.1$--$1.4$ (see, \eg, \citealp{Metzger:2010sy,Korobkin:2012uy}).
Investigations of the \ac{KN} AT2017gfo found a good agreement of the light curve with radioactive heating from $r$-process nuclei \citep{Tanaka:2017qxi,Waxman:2017sqv,Kasliwal:2017ngb,Murguia-Berthier:2017kkn,Cowperthwaite:2017dyu,Wu:2018mvg}.
This is broadly consistent with simulations that find dynamical ejecta are typically at least mildly neutron rich \citep{Rosswog:1998hy,Hotokezaka:2013iia,Radice:2018pdn,Nedora:2020hxc,Kiuchi:2022nin}, although the ejected mass is insufficient to explain the luminosity and no ab initio simulation can yet quantitatively reproduce the light curves.

Tens to hundreds of milliseconds after the merger, spiral-wave \citep{Nedora:2019jhl,Nedora:2020hxc} and neutrino-driven winds \citep{Dessart:2008zd,Perego:2014fma,Martin:2015hxa,Fujibayashi:2020dvr} can launch $\gtrsim 10^{-3} M_\odot$ of ejecta from the remnant accretion disk as long as the remnant does not collapse to a \ac{BH} shortly after the merger.
Recent long-term \ac{BNS} merger simulations presented in \citet{Just:2023wtj,Radice:2023zlw,Schianchi:2023uky,Bernuzzi:2024mfx,Neuweiler:2025klw} have found that a large fraction of the neutrino-driven wind exhibits electron fractions close to 0.5.
These simulations employ a two-moment ``M1'' neutrino-transport scheme, unlike previous long-time simulations of \ac{BNS} mergers, which typically use a more approximate leakage-based neutrino transport \citep{Nedora:2020hxc,Combi:2022nhg,Kiuchi:2022nin}.
As a consequence of the relatively high electron fraction, the nucleosynthesis in the neutrino-driven wind is dominated by an $\alpha$-rich freeze-out and by the production of \nic.
In contrast to $r$-process powered transients, the heating rate in the proton-rich wind ejecta follows the exponential decay of \nic and \cob with half-lives of 6 and 77 days, respectively.
If a large enough amount of \nic is produced in the neutrino-driven ejecta from \ac{BNS} mergers, this could observably alter the associated \ac{KN} light curve compared to what would be expected from purely neutron-rich ejecta.
\citet{Just:2023wtj} find that iron-group elements are produced in the neutrino-driven wind.
However, they do not report a significant enhancement of the \ac{KN} luminosity.
While it was confirmed that the decay of \nic alone cannot account for the transient observed in 2017 \citep{Cowperthwaite:2017dyu}, to our knowledge it has not been investigated whether the combined heating from \nic and $r$-process elements are compatible with AT2017gfo.

In this work, we investigate the production of \nic in neutrino-driven ejecta from \ac{BNS} mergers and the impact on the \ac{KN} emission based on the simulations presented in \citet{Radice:2023zlw,Gutierrez:2024pch,Bernuzzi:2024mfx}.

\section{Methods}
We analyze and compare the ejection of proton-rich material in four \ac{NR} \ac{BNS} merger simulations summarized in \cref{tab:1}.
\begin{table*}[t]
  \centering
  \caption{\label{tab:1}%
    Overview of the \ac{BNS} models considered in this work.
    Given are initial gravitational \ac{BNS} masses $M_1$ and $M_2$, the simulated postmerger time $t_{\text{pm}}$, the time until \ac{BH} formation $t_{\text{BH}}$,
    the total ejecta mass $M_{\text{ej}}$, the mass of very neutron-rich ($Y_e < 0.25$) ejecta $M_{\text{ej}}^{n}$, and proton-rich ($Y_e > 0.5$) ejecta $M_{\text{ej}}^{p}$ and the total mass of ejected He and \nic $M_{\text{ej}}^{\text{He}}$, $M_{\text{ej}}^{\text{Ni}}$, respectively.
    Ejecta masses are inferred using the Bernoulli criterion ($-h u_t > 1$).
  }%
  \begin{ruledtabular}
  \begin{tabular}{l|rr|rr|rrrrr}
    Model &
    $M_1$ &
    $M_2$ &
    $t_{\text{pm}}$ &
    $t_{\text{BH}}$ &
    $M_{\text{ej}}$ &
    $M_{\text{ej}}^{n}$ &
    $M_{\text{ej}}^{p}$ &
    $M_{\text{ej}}^{\text{He}}$ &
    $M_{\text{ej}}^{\text{Ni}}$ \\
    &
    \multicolumn{2}{c|}{[$M_\odot$]} &
    \multicolumn{2}{c|}{[ms]}  &
    \multicolumn{5}{c}{[$10^{-3} M_\odot$]} \\
    \hline
    \BLh  & 1.64 & 1.15 & 103 & 114    & 10.85 &  4.01 & 3.72 & 3.03 &  0.84 \\
    \DDa  & 1.81 & 1.08 & 111 & $>$130 & 25.31 & 19.30 & 1.59 & 1.56 &  0.43 \\
    \DDs  & 1.35 & 1.35 &  79 & $>$79  &  5.80 &  1.18 & 1.53 & 1.23 &  0.41 \\
    \SFH  & 1.35 & 1.35 &  35 & 8.3    &  6.25 &  3.51 & 0.00 & 0.09 &  0.01
  \end{tabular}
  \end{ruledtabular}
\end{table*}
They consist of two asymmetric ($q>1$), \BLh and \DDa, and two symmetric ($q=1$) \ac{BNS} systems, \SFH and \DDs, employing three different microphysical \acp{EOS}: SFHo \citep{Steiner:2012rk}, DD2 \citep{Typel:2009sy,Hempel:2009mc}, and BLh \citep{Bombaci:2018ksa,Logoteta:2020yxf}.
All models discussed in this work have been presented in previous publications \citep{Radice:2023zlw,Gutierrez:2024pch,Bernuzzi:2024mfx}.
Note that we use the baryon-mass ratio $q = M_1^{\text{b}}/M_2^{\text{b}} \geq 1$ for the naming of models to stay consistent with the naming convention in \citet{Bernuzzi:2024mfx} while \cref{tab:1} lists the gravitational masses $M_1, M_2$.

All simulations are performed with the \thc~code~\citep{Radice:2012cu,Radice:2013hxh,Radice:2015nva,Radice:2013xpa,Radice:2016dwd,Radice:2021jtw}, which is built on top of the \texttt{Cactus} framework \citep{Goodale:2003a,Schnetter:2007rb}.
The simulations employ a truncated, two-moment gray neutrino-transport scheme that retains all nonlinear neutrino-matter coupling terms~\citep{Radice:2021jtw} and uses the Minerbo closure.
The set of weak reactions included in the transport scheme is described in, \eg, \citet{Galeazzi:2013mia,Radice:2016dwd,Perego:2019adq}.
The general-relativistic hydrodynamics equations are formulated in conservative form (see \citealp{Radice:2018pdn} for details on the precise equations solved here) and augmented by a large-eddy scheme that accounts for angular momentum transport due to magnetohydrodynamical effects \citep{Radice:2017zta, Radice:2020ids, Radice:2023zlw}.

The equal-mass simulations, \DDs and \SFH, consist of two $1.35 M_\odot$ \acp{NS}.
In the \BLh model, the \ac{BNS} masses are compatible with the chirp mass measured in GW170817, and the mass ratio lies at the upper bound of the constraints inferred with low spin priors \citep{Abbott:2018wiz}.
The \DDa model exhibits an even larger mass ratio and as such, represents a comparably extreme case.
Of the four binaries considered in this work, only the remnant in \SFH collapses to a \ac{BH} in the gravitational-wave driven phase, while the central remnant is a massive \ac{NS} for the duration of the simulation in the other three models.
These models eject matter for the full duration of the simulations until $\sim \unit[100]{ms}$ post merger.

After the time at which the ejecta analyzed in this work were extracted, we continued the simulation of the models \BLh and \DDa.
In \BLh, the merger remnant collapses after \unit[114]{ms} postmerger (\ie, shortly after the time at which the analysis in this Letter is performed).
The simulation of \DDa was performed until $\sim$\unit[130]{ms} post merger without the remnant collapsing to a \ac{BH}.
At this point, the evolution of the gauge becomes unstable, and the simulation crashes.
The model \DDs is studied in many works in the literature, including \citet{Fujibayashi:2020dvr}, who evolved the postmerger remnant in axisymmetry for several seconds without it collapsing.

  We do not expect that further evolving this model will result in a qualitative change of the ejecta properties unless it is evolved to viscous timescales, which is not computationally feasible with our 3D M1 GRMHD simulations.
  The ejection of matter stops shortly after collapse in the model \SFH, so we do not continue the simulation further.

We evolve the ejecta and the nucleosynthesis during the expansion up to $t=\unit[100]{days}$ with the Lagrangian radiation-hydrodynamics code \snec \citep{Morozova:2015bla,Wu:2021ibi} coupled to the \skynet nuclear network \citep{Lippuner:2017tyn}, as described in \citet{Magistrelli:2024zmk,Magistrelli:2025xja}.

Ejecta properties are recorded on the extraction sphere as a function of time and the azimuthal and polar angles.
The data are then averaged over the azimuthal angle, weighted by the density as described in \citet{Magistrelli:2024zmk}.
Finally, we create initial Lagrangian profiles for SNEC with the layered shell-by-shell method described in \citet{Wu:2021ibi}.
The dependency on the polar angle of the ejecta properties is approximately included in a ray-by-ray fashion by mapping the initial profile into an effective 1D problem for a number of angular bins.
One full simulation includes 26 rays uniformly spaced in the polar angle, ranging from $\theta=0$ (polar) to $\theta=\pi/2$ (equatorial), assuming mirror symmetry across the equatorial plane.
To ensure that all the intensive quantities (including the density) remain unaltered, we multiply the total mass by the scaling factor $\lambda_\theta = 4\pi / \Delta\Omega$, where $\Delta\Omega \simeq 2\pi \sin{\theta} \, d\theta$ is the solid angle of the angular bin.
Each angular section is discretized radially into 1000 fluid elements and evolved independently up to $t=\unit[100]{days}$.
Nonradial fluid motion is neglected in this approach, which we expect to slightly change the angular distribution of the ejecta and therefore, the observer-angle dependence of the light curve.
However, this effect is most likely negligible compared to uncertainties in the \acp{EOS} and the numerical simulations.

We note that the ejecta velocities change by up to a few percent of the speed of light after the transition from the \thc to the \snec simulations.
This is likely caused by the method of initialization of the \snec simulations (see \citealp{Wu:2021ibi}), the transition from a general relativistic to a Newtonian description of gravity, and the switch of the employed \acp{EOS}.
Because the regions that constitute \nic are unaffected, the conclusions of this Letter remain unchanged.

The nuclear network includes 7836 isotopes up to~$^{337}$Cn and uses the JINA REACLIB \citep{Wiescher:2010eni} and the same setup as in \citet{Lippuner:2015gwa, Perego:2020evn}.
We self-consistently couple the effects of nuclear heating from the in situ network with the hydrodynamics evolution.
The thermalization of energy produced by nuclear reactions is treated explicitly for contributions from $\gamma$-rays, $\alpha$ particles, electrons, and other nuclear reactions products and is described in detail in \citet{Magistrelli:2024zmk}.
For some ejecta components, the temperature at the moment of mapping has already dropped below $\sim \unit[6]{GK}$.
To ensure consistent initialization in \ac{NSE} conditions, we extrapolate to earlier times using an analytic trajectory starting at $T=\unit[8]{GK}$ as described in \citet{Magistrelli:2025xja}.
The details of the extrapolation change the \nic abundance by $\sim 10\%$ (see Appendix B in \citet{Magistrelli:2025xja}).
To calculate the luminosity in each angular section, we employ the analytic, time-independent opacity described in \citet{Wu:2021ibi}.
We note that \citet{Magistrelli:2025xja} investigate the impact of composition and time-dependent opacities and find that the overall shape of the light curve is impacted.

The \ac{KN} light curves are combined depending on the viewing angle as described in \citet{Martin:2015hxa,Perego:2017wtu} to approximately take nonradial photon motion into account.
Note that this recombination assumes spherical symmetry, which, given the asymmetry of the optical depth in the ejecta, might not be fully accurate.
Despite these approximations, we do not expect the conclusions drawn in this Letter to be affected.

KN spectra from each angular section are computed using the radiative transfer spectral synthesis code \tardis \citep{Kerzendorf:2014tardis}.
The density, velocity, and composition profiles are obtained by rebinning the fluid elements outside the photosphere into 50 radial shells, while the input luminosity is given by the bolometric luminosity, $L = \sum_\theta L_\theta / \lambda_\theta$ where $L_\theta$ is the luminosity from the individual ray.
We use the atomic dataset developed by \citet{Gillanders:2022opm}, considering atomic lines of elements with $Z\le40$ (the first nine rows in Table 1 of \citet{Gillanders:2022opm}).
Since we are mostly interested in modeling the polar ejecta, which predominantly synthesize light $r$-process elements, this limited line list is likely sufficient.
Ionization and excitation are treated with the local thermal equilibrium and dilute local thermal equilibrium approximations, respectively, while the ``macroatom'' scheme is employed for line interaction treatment.
In addition to KN spectra obtained from each ray, we combine them for the fixed viewing angle, following the same weighting prescription used for the luminosity computed in the \snec simulations.

\section{Results}
\subsection{Ejecta Distribution and Properties}
\begin{figure}[t]
    \includegraphics[width=\columnwidth]{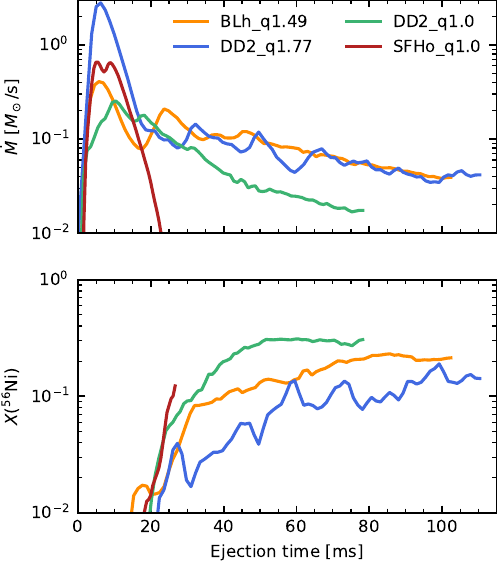}
  \caption{\label{fig:ejection_rate}%
    Top panel: mass-ejection rate for the considered \ac{BNS} models.
    Bottom panel: \nic mass fraction that the fluid elements will achieve after \unit[10]{s} as a function of their ejection time.
    The $x$-axis shows the time of ejection defined as the time at which the fluid elements cross the sphere at $\sim \unit[450]{km}$.
          }
\end{figure}
The upper panel of \cref{fig:ejection_rate} shows the mass-ejection rate as a function of ejection time for the four models considered in this work.
Here ``ejection time'' refers to the time at which the Lagrangian fluid elements cross the sphere at $\sim \unit[450]{km}$ and is therefore slightly delayed with respect to the actual time of ejection from the merger remnant.
All models exhibit dynamical ejecta of tidal and shock-heated origin within the first \unit[10]{ms} after merger.
After that, oscillations of the massive \ac{NS} drive a spiral-wave wind that ejects moderately neutron-rich matter, leading to an oscillatory mass-ejection rate.
This ejection continues until the end of the simulation or the collapse of the remnant.
Such a spiral-wave wind is stronger in the asymmetric models, especially in \DDa, where it persists until the end of the simulation.

The neutrino-driven mass ejection starts $\sim \unit[10]{ms}$ after the merger in the polar direction ($\theta \lesssim 30^\circ$) and continues until the end of the simulation or the collapse of the remnant.
In the almost exclusively proton-rich polar ejecta, the freeze-out from \ac{NSE} is dominated by $\alpha$ particles, resulting in a final He mass fraction of $\sim 10\%$, in agreement with the values found in \citet{Just:2023wtj}.
This could lead to a detectable feature in the \ac{KN} spectrum after a few days \citep{Perego:2020evn,Tarumi:2023apl,Sneppen:2024ros}, indicating the presence of a long-lived \ac{NS} remnant \citep{Sneppen:2024jch}.

Furthermore, $\alpha$ captures synthesize nuclei along the $Z=N$ line on the nuclear chart ending at \nic due to its doubly magic nature.
The lower panel of \cref{fig:ejection_rate} shows the \nic mass fraction of the fluid elements as a function of the time at which they cross the extraction sphere.
The \nic fraction $X(^{56}\mathrm{Ni})$ refers to the value at \unit[10]{s} for each fluid element because \nic is not yet fully synthesized at $r=\unit[450]{km}$.
In the noncollapsing models, the mass fraction of \nic reaches $\sim 10\%$--$30\%$ and increases with time as the neutrino-driven wind becomes more proton rich.
\DDa exhibits a comparably low mass fraction due to its long-lasting spiral-wave wind ejecta component, which ejects slightly neutron-rich matter alongside the polar proton-rich wind, thereby lowering the relative fraction of \nic.
We also observe the beginning of a neutrino-driven wind and the production of \nic in \SFH, but due to its early collapse, the overall mass is much smaller than in the noncollapsing models and therefore negligible.

Note that a neutrino-driven wind is also present in previous \ac{BNS} merger simulations using the leakage scheme and ``M0'' neutrino absorption scheme \citep{Radice:2018pdn,Nedora:2020hxc,Combi:2022nhg}.
The crucial difference is that the M1 neutrino-transport scheme leads to a more massive and more proton-rich wind.
In the M0 neutrino scheme, neutrinos are only transported in a radial ray-by-ray fashion.
The region with the highest neutrino emissivity, however, lies at the remnant--disk interface (\ie, off-center).
Thus, the neutrino-driven wind is not as massive and is ejected closer to the equatorial plane in M0 simulations.
In the M1 scheme, however, neutrinos emitted from the disk interface can move vertically, leading to a more massive neutrino-driven wind with a more polar ejection angle, as already visible in older simulations that did not prescribe radial neutrino fluxes \citep{Perego:2014fma}.

\begin{figure}[t]
  \centering
  \includegraphics[width=\columnwidth]{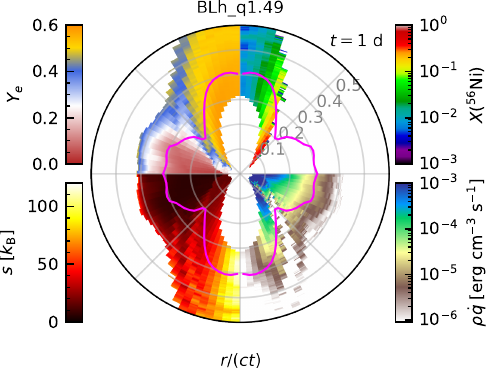}
  \caption{\label{fig:grid_plots}%
    Spatial distribution of the ejecta of the \BLh model after 1 day.
    Top-left quadrant: electron fraction at the beginning of the \snec simulation.
    Top-right quadrant: mass fraction of \nic.
    Bottom-left quadrant: initial entropy at the beginning of the \snec simulation.
    Bottom-right quadrant: local heating rate.
    The magenta line indicates the position of the photosphere.
    Regions with $Y_e \gtrsim 0.48$ and $s \lesssim 50 k_\mathrm{B}/$baryon produce a large amount of \nic, which has a significant impact on the heating rate on the timescale of days.
          }
\end{figure}
\Cref{fig:grid_plots} shows the spatial distribution of the ejecta 1 day after merger for the \BLh model.
The top and bottom left quadrants show the spatial distribution of the initial electron fraction and entropy, respectively, as extracted from the NR simulation.
The tidal ejecta are located close to the equatorial plane with low electron fractions and entropies, while the shock-heated ejecta are more spread out and exhibit a large range in electron fraction and entropy.
In the asymmetric models, the tidal ejecta are more massive than the shock-heated ejecta (in \DDa, the tidal ejecta contribute $\sim 50\%$ of the total ejecta mass), while the shock heated ejecta are more important in the equal-mass models.

\subsection{Nucleosynthesis}
If the initial electron fraction of the ejecta is even slightly neutron rich, the small amount of free neutrons drive the composition away from the $Z=N$ line and toward the valley of stability.
To produce \nic, the electron fraction must therefore be fairly high ($Y_e\gtrsim 0.48$).
This condition leads to a sharply bounded region where the \nic mass fractions lie between 1\% and 30\%, as visible in the top-right quadrant in \cref{fig:grid_plots} showing the mass fraction of \nic after 1 day.
Higher entropies ($s \gtrsim 50 k_{\text{B}}$) disfavor the formation of seed nuclei from $\alpha$ particles, leading to less effective production of \nic and a higher He mass fraction.
Therefore, the \nic mass fraction decreases for  angles close to the polar axis.
The high-entropy ejecta could potentially host the neutrino p ($\nu$p) process \citep{Pruet:2005qd,Frohlich:2005ys,Wanajo:2006ec,Wanajo:2017cyq,Nishimura:2019jlh}.
In this case, electron-antineutrino absorption on free protons create free neutrons, which are immediately captured by proton-rich seed nuclei.
This way, elements between the iron group and Ag including light $p$-nuclei can be produced.
However, our network calculations do not include neutrino capture reactions and can therefore not capture this mechanism.
The impact on the \nic abundance and the conclusions drawn in this Letter is likely negligible.

The overall ejecta masses, the mass of ejecta hosting a strong $r$-process $M_\text{ej}^r$ (which we approximate as the ejecta with $Y_e < 0.25$), proton-rich ejecta $M_\text{ej}^p$ ($Y_e > 0.5$), He, and \nic are reported in \cref{tab:1}.
The mass of proton-rich ejecta exceeds $10^{-3} M_\odot$ in all noncollapsing models and is highest in \BLh.
In the proton-rich ejecta, the \nic mass fraction is approximately $\sim 22\%$--$27\%$.
At the end of the simulations employing the DD2 \ac{EOS}, the neutrino-driven wind is still ejecting matter with a high electron fraction.
The \nic masses reported in \cref{tab:1} as well as the results that follow from it should therefore be considered lower limits in these models.

\subsection{Kilonova Light Curve}
The lower right quadrant in \cref{fig:grid_plots} shows the local heating rate at $t = \unit[1]{day}$ in the \BLh model.
There are two distinct regions visible.
In the neutron-rich equatorial region, the decay of $r$-process material dominates, while the decay of \nic is almost exclusively responsible for the nuclear energy generation in the polar region.
At 1 day (roughly when the photosphere starts to enter the bulk of the ejecta), the energy generation in the two regions is already comparable.
\begin{figure*}[t]
  \centering
  \includegraphics[width=\textwidth]{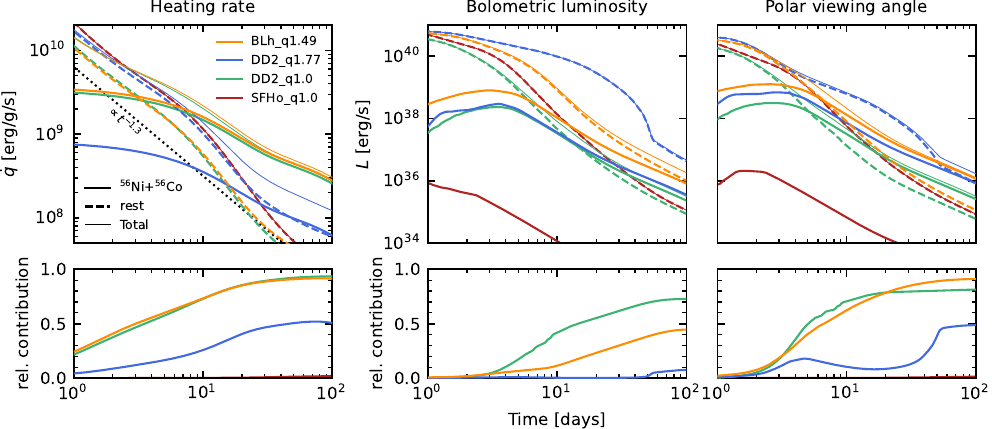}
  \caption{\label{fig:heating}%
    Mass-averaged nuclear energy generation (left panels), bolometric luminosity (middle panels) and luminosity observed at a viewing $20^\circ$ (right panels).
    The upper panels show the heating and luminosities generated by the decay of \nic and \cob (solid lines), the contribution of all remaining isotopes (dashed lines), and the total (thin lines).
    The dotted black line in the upper left panel indicates a heating rate proportional to $t^{-1.3}$.
    The lower panels show the relative contribution of \nic and \cob in each of the above panels.
    }
\end{figure*}
\Cref{fig:heating} shows nuclear energy generation and luminosities for the four simulations considered in this work.
Solid lines in the left panel show the mass-averaged specific energy generation $\dot{q}$ due to the decay of \nic and \cob, dashed lines show the combined contribution of all other isotopes, and thin lines show the total.
In \BLh and \DDs, the energy generated by the decay of \nic makes up more than 50\% of the total heating after 3 days and more than 75\% after 10 days.

Due to the large tidal ejecta mass in \DDa the relative contribution of \nic is reduced, resulting in 10 (25)\% after 3 (10) days.
In all three models, the decay of \cob makes up the largest part of the total energy generated around 100 days.
Since the heating rate follows the decay of two nuclei, it deviates  significantly from the power-law behavior $\propto t^{-1.3}$ (shown as a dotted black line) expected for the combined heating of $r$-process elements.
The combined heating from all other decays (dashed lines) is reasonably well described by the power law.
The slight bump at a few days is caused by the $\beta$ decay of $^{132}$I, which is fed by the decay of $^{132}$Te with a half-life of \unit[3.2]{days} \citep{Kullmann:2021gvo}.

To estimate their contribution to the light curve, we separately calculate the thermalized heating rate of \nic and \cob outside the photosphere.
The middle panel of \cref{fig:heating} shows the bolometric luminosity for each of the simulations (solid lines).
Similar to the left panel, solid lines show the contribution from the decay chain of \nic outside the photosphere and dashed lines show the difference between the total luminosity (thin lines) and the \nic and \cob contribution.
Within the first 5 days, the \nic is partially hidden within the photosphere, and its impact is thus not yet visible.
Furthermore, the impact of the \nic heating on the bolometric light curve is smaller compared to its impact on the nuclear energy generation due to the fact that the energy of the decay is exclusively released in $\gamma$-rays and neutrinos.
Since the neutrino-driven ejecta become mostly transparent to $\gamma$-rays within the first 5--10 days, the energy generated by \nic does not thermalize effectively.
In the equatorial region, on the other hand, the $\gamma$-ray opacities are higher and a large fraction of the energy generated by $r$-process nuclei is released by $\beta$-decay electrons which thermalize more efficiently.
While \cob also decays primarily via electron capture, it can also decay via $\beta^+$ decay, emitting on average \unit[121]{keV} via positrons per decay \citep{Wu:2018mvg}.
Therefore, after $t \gtrsim \unit[10]{days}$, the impact of \cob on the bolometric luminosity increases significantly.
In $\DDa$, where the ratio of the $r$-process enriched to \nic enriched ejecta is very high (see $M_{\text{ej}}^{n}$ versus $M_{\text{ej}}^{p}$ in \cref{tab:1}), the heating from \nic and \cob\ contributes less than 10\% to the total bolometric luminosity.

For a polar viewing angle, the neutrino-driven ejecta fill a large fraction of the field of view while the tidal ejecta are subdominant.
Consequently, the contribution of \nic and \cob to the observed luminosities is significantly enhanced and reaches 50\% after 5--7 days for \DDs and \BLh.
In \DDa, the contribution of \nic and \cob to the observed luminosity is still mostly overshadowed by the decay of $r$-process elements where the most relevant isotopes are $^{132}$I and later, the $\alpha$ emitters $^{225}$Ac, which decays with a half-life of 10 days \citep{Kullmann:2021gvo} and subsequently $^{221}$Fr.
After $t \gtrsim \unit[50]{days}$, their decay becomes less important, and at the same time, $\alpha$ particles begin to thermalize less efficiently than electrons and positrons.
Thus, at these late times the decay of \cob becomes more relevant again.

We calculate the EM spectra after $t=\unit[1]{day}$ for \BLh with \tardis using the ray-by-ray approach outlined above.
The two most prominent features in the spectra are created by Sr and Ca \citep{Watson:2019xjv,Domoto:2021xfq,Perego:2020evn}.
Ca is primarily produced in the form of $^{48}$Ca in the slightly neutron-rich part of the ejecta, but $\sim 10\%$ is produced as $^{40}$Ca in the proton-rich wind.
Iron-group elements are not visible in the spectra due to the low luminosities at the relevant wavelengths and the dominance of Ca and Sr lines. 
We thus cannot identify a direct indication of proton-rich ejecta in the early spectra.
While this result disfavors the emergence of an "iron-like" feature as suggested by the photometric analysis of \citet{Jin:2016pnm} for the afterglow of GRB 050709, it supports instead modeling efforts for the AT2017gfo spectra around the chosen epoch (see, e.g., \citet{Domoto:2021xfq,Domoto:2022cqp,Gillanders:2022opm}).

The presence of r-process elements throughout the angle and velocity distribution of the ejecta shifts the underlying blackbody spectrum to higher wavelengths, as a result of the high line opacities of such species in the optical/UV.
The latter translates in a strong line blanketing redistributing photons to lower energies, and, eventually, in a cooler blackbody due to the longer photon diffusion time (see, e.g., \citet{Kasen:2013xka,Tanaka:2013ana}).

Note that the included line-list (only Zr and lighter elements) is not sufficient to produce an accurate spectrum.
However, we only perform this simple spectral modeling with a limited line list to gauge if the large abundance of Ni can produce a visible feature in the spectrum during the photospheric phase.
Since we do not find any such feature, we do not pursue more realistic spectral modeling because it likely does not change the conclusion above and is outside the scope of this work.

\subsection{Gamma-Ray Spectra}
Most of the $\gamma$-rays emitted by the decay of \nic and \cob escape the ejecta and thus contribute to the $\gamma$-ray spectrum.
Previous works have investigated the possibility of direct detection of $\gamma$-rays from the decay of $r$-process nuclei \citep{Hotokezaka:2015cma,Li:2018wee,Korobkin:2019uxw}, however, they did not include the emission from iron-group elements.
We calculate the $\gamma$-ray emission spectrum at $t = \unit[7]{days}$ for \BLh accounting for the relativistic Doppler shift for each fluid element individually.
\begin{figure}[t]
  \centering
  \includegraphics[width=\columnwidth]{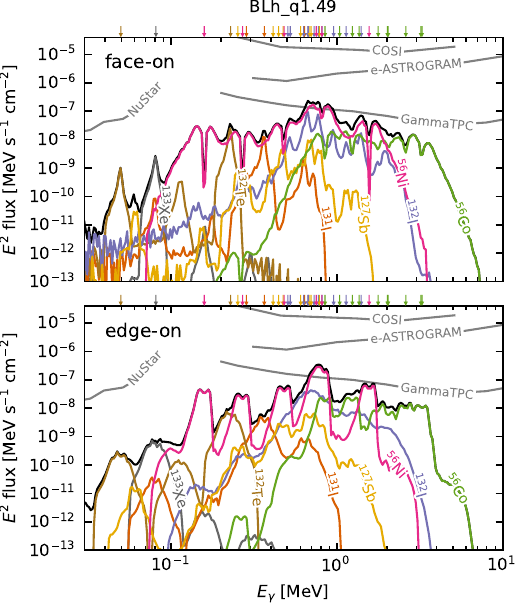}
  \caption{\label{fig:gam_spec_7d}%
    $\gamma$-ray emission spectrum at $t = \unit[7]{days}$ for \BLh at \unit[40]{Mpc} for a polar (upper panel) and an equatorial (lower panel) viewing angle.
    Colored lines and arrows on top of the panels indicate the contribution of individual isotopes and the original line positions, respectively.
    Gray lines show the sensitivity of the NuStar, COSI, e-ASTROGRAM, and GammaTPC telescopes.}
\end{figure}
The resulting spectrum is shown in \cref{fig:gam_spec_7d} for a polar (upper panel) and an equatorial viewing angle (lower panel), assuming an event distance of \unit[40]{Mpc}.
The partial spectra from the most relevant isotopes are shown by colored lines and the position of the corresponding unshifted emission lines are indicated by arrows above the panels.
In the energy range between \unit[100]{keV} and \unit[10]{MeV}, both spectra are dominated by \nic and \cob lines.
From a polar viewing angle, the wind ejecta are moving either away (below the merger remnant) or toward the observer (above the merger remnant).
Lines from \nic and \cob are thus either red- or blueshifted and form a double peak.
$\gamma$-rays from $r$-process elements stem mostly from the tidal ejecta in the equatorial plane, resulting in narrower lines.
For an edge-on equatorial observer, peaks from \nic and \cob are instead narrower, while those from $r$-process isotopes are broader.
Before $\sim 1$ day, the $\gamma$-rays produced by the decay of \nic do not escape the ejecta, and the \nic peak in the spectrum is not yet visible.
At $\sim 3$--$15$ days, the \nic lines are the most dominant peaks in the spectrum.
After $\sim 15$ days, most \nic has decayed, and \cob features are the most dominant.

We also include the sensitivity curve of NuStar, the Compton Spectrometer and Imager (COSI) \citep{Tomsick:2023aue}, e-ASTROGRAM \citep{e-ASTROGAM:2017pxr}, and GammaTPC \citep{Shutt:2025xvc}.
Even though $\gamma$-ray emissions from \nic and \cob increase the peak flux by an order of magnitude, the detection with COSI would require an event within the very optimistic distance of $\sim \unit[5]{Mpc}$, while e-ASTROGRAM would require $\sim \unit[15]{Mpc}$.
However, GammaTPC would be sensitive enough to detect the double peak due to the decay of \nic around \unit[700--800]{keV} for an event at a distance of \unit[40]{Mpc}.

\section{Discussion}
We analyzed the outflow of four long-term \ac{BNS} merger simulations employing an M1 neutrino-transport scheme.
All noncollapsing models eject large amounts of proton-rich ejecta ($\sim 10^{-3} M_\odot$) synthesizing $\sim 4$--$8 \times 10^{-4} M_\odot$ of \nic, the decay of which can significantly alter the late ($t \gtrsim 5\,\mathrm{days}$) light curve.
The observation of this effect in a \ac{KN} would strongly indicate that the remnant of the \ac{BNS} merger did not collapse to a \ac{BH} for at least $\sim$ \unit[100]{ms} after the merger.

For a confirmation of our results, more work will be needed in the future.
We discuss the main uncertainties in our work below.

More magnetohydrodynamics simulations including a sophisticated neutrino-transport scheme will be necessary, since magnetic fields might launch magnetically driven winds on similar timescales to the neutrino-driven wind \citep{Kiuchi:2022nin,Combi:2022nhg}.
This could increase the amount of polar ejecta but might reduce their electron fraction by reducing the expansion timescale of the wind ejecta and thus the total amount of neutrino-captures per escaping fluid element.
Note, however, that a wound-up magnetic field close to the massive \ac{NS}'s polar cap could trap fluid elements for longer and therefore increase the time of neutrino irradiation, similar to what is observed in magnetorotational supernova simulations \citep{Obergaulinger:2020cqq}.
The recent work by \citet{Musolino:2024sju} presents \ac{BNS} merger simulations including both an M1 neutrino transport and magnetic fields.
While their figure 2 indicates the presence of some proton-rich ejecta, they do not provide quantitative information about the composition of the ejected material.
The GRMHD simulations with an M1 neutrino scheme presented in \citet{Neuweiler:2025klw} also show a clear enhancement in proton-rich ejecta.
However, their tracer-particle scheme does not resolve this outflow, and the production of \nic can thus not be analyzed in detail.
In preliminary MHD M1 simulation obtained with the \texttt{GR-Athena++} code \citep{Daszuta:2021ecf,Cook:2023bag}, we find an enhancement in proton-rich ejecta if magnetic fields are included, but further analysis is needed for any definitive conclusions.

M1-type schemes are known to produce an unphysical overabundance of neutrinos in the polar regions \citep{Frank:2007,Foucart:2020qjb,Gizzi:2021ssk,Foucart:2024npn}, so one could conclude this to be responsible for the strong neutrino-driven wind in our simulations.
However, \citet{Radice:2021jtw} found that the THC M1 neutrino-transport also produces a strong proton-rich wind with the Eddington closure which preserves the linearity of the transport operator and is therefore not affected by radiation shocks.
\citet{Radice:2021jtw} thus conclude that the proton-rich wind is not the result of an unphysical ``beam-crossing'' artifact.
Furthermore, \citet{Foucart:2024npn} found good agreement between the M1 scheme employed in our simulations and Monte Carlo transport schemes.

The neutrino transport employed in our 3D simulations, while being state-of-the-art for long simulations, is only approximate.
Future studies with an energy-dependent neutrino-transport scheme and more accurate set of weak reactions will be needed to confirm the presence of proton-rich neutrino-driven winds from \ac{BNS} merger remnants.
\Citet{Cheong:2024buu} perform axisymmetric MHD simulations of postmerger-like hypermassive \acp{NS} in the conformal-flatness approximation and compare energy-integrated to energy-dependent M1 neutrino schemes.
They find that the energy-integrated prescription tends to overestimate the electron fraction of the ejecta within the first \unit[50]{ms} of postmerger evolution.
However, \citet{Just:2023wtj} also employ an energy-dependent M1 scheme and find significant proton-rich ejecta.
Future three-dimensional simulation in full GR will likely be needed to accurately gauge the impact of energy-dependent M1 schemes.

The simulations presented in this work only span $\sim \unit[100]{ms}$ of the postmerger phase.
Simulations in axisymmetry have shown that the ejection of matter might continue for seconds due to viscous heating caused by the magnetorotational instability \citep{Fujibayashi:2020dvr,Fujibayashi:2022ftg,Just:2023wtj}.
This ejection mechanism can occur for \ac{BH} and massive \ac{NS} remnants and might dilute previously ejected proton-rich with slightly neutron-rich matter.
In this case, the signature of the decay of \nic and \cob might be hidden.
However, the details of this very late phase of matter ejection are highly uncertain because second-long simulations are currently only possible with very approximate, axisymmetric viscous hydrodynamics simulation setups.

While the presence of iron-group elements has a clear impact on the nuclear energy generation, uncertainties in the thermalization efficiency complicate the quantitative characterization of the effect on the light curve.
Since the thermalization efficiency of the generated energy is expected to drop rapidly on the timescale of a few days, the light curve depends sensitively on the employed thermalization treatment.
This time frame overlaps with the decay time of \nic and \cob which complicates the prediction of the expected feature.
The light-curve modeling employed in this work is fairly simple compared to more sophisticated radiative transport codes.
The impact of \nic and \cob heating on the light curve predominantly stems from a region in the ejecta that is mostly transparent on the relevant timescales of days.
Thus, we expect the main source of uncertainty during the light-curve modeling to stem from the thermalization and not the radiative transport.

Note that most of these uncertainties could decrease or increase the impact of \nic and \cob on the \ac{KN}.
Thus, we draw the following conclusion:
On one hand, if a future \ac{KN} observation matches the predicted signature of the decay of \nic and \cob, this strongly indicates the presence of a long-lived ($\gtrsim \unit[100]{ms}$) merger remnant.
On the other hand, if no such match is found, this could be either due to an early collapse or because proton-rich ejecta do not significantly impact the electromagnetic signal due to any of the abovementioned caveats.
In this case, an interpretation requires more detailed modeling in the future to draw a definitive conclusion.

In the meantime, we suggest that future parametric studies of \acp{KN} should include the possibility for the presence of \nic in the ejecta.
By systematically varying the abundance of \nic among other parameters such as ejecta mass and velocity, in a large-scale parametric study, it could be determined if the light curves of AT2017gfo and future \ac{KN} observations are compatible with the presence of \nic independently of the simulation uncertainties mentioned above.

Furthermore, a large abundance of iron-group elements could be verified in the nebular-phase spectra.
Given the strong asymmetry in the ejecta composition, the very high relative amount of \nic ($\sim20\%$--$30\%$) in the proton-rich ejecta could lead to the presence of strong Ni and Co features on a timescale of weeks, possibly in the wavelength range $\sim7000$--$7500\,$\AA\ \citep{Maguire:2018vtv} or in the infrared \citep{Pognan:2025eat,Jerkstrand:2025rjr}.
However, a consistent modeling of the ionization structure would be needed in order to predict the overall and relative strengths of such features \citep{Pognan:2023qhw}.

\begin{acknowledgments}
We would like to thank David Radice for providing simulation data, James Gillanders for providing atomic data for \tardis and Oliver Just, Kenta Hotokezaka, and Moritz Reichert for fruitful discussions and comments.
S.B.\ and M.J.\ acknowledge support by the EU Horizon under ERC Consolidator Grant, no.\ InspiReM-101043372.
F.M.\ acknowledges support from the Deutsche Forschungsgemeinschaft (DFG) under grant No.\ 406116891 within the Research Training Group RTG 2522/1.
G.R.\ and A.A.\ acknowledge support from the Deutsche Forschungsgemeinschaft (DFG, German Research Foundation, Project ID 279384907, SFB 1245) and the State of Hesse within the Research Cluster ELEMENTS (Project ID 500/10.006).
E.L.\ acknowledges support by the European Union – NextGenerationEU RFF M4C2 1.1 PRIN 2022 project 2022RJLWHN URKA.
The work of A.P.\ is partially funded by the European Union - Next Generation EU, Mission 4 Component 2 - CUP E53D23002090006 (PRIN 2022 Prot.\ No.\ 2022KX2Z3B).
Simulations were performed on SuperMUC-NG at the Leibniz-Rechenzentrum (LRZ) Munich and on the national HPE Apollo Hawk at the High Performance Computing Center Stuttgart (HLRS).
The authors acknowledge the Gauss Centre for Supercomputing e.V.\ (\url{www.gauss-centre.eu}) for funding this project by providing computing time on the GCS Supercomputer SuperMUC-NG at LRZ (allocations {\tt pn36ge}, {\tt pn36jo}, {\tt pn68wi}, and {\tt pn76li}).
The authors acknowledge HLRS for funding this project by providing access to the supercomputer HPE Apollo Hawk under the grant Nos.\ INTRHYGUE/44215 and MAGNETIST/44288.
Postprocessing and development runs were performed on the ARA cluster at Friedrich Schiller University Jena.
The ARA cluster is funded in part by DFG grants INST 275/334-1 FUGG and INST 275/363-1 FUGG, and ERC Starting Grant, grant agreement No.\ BinGraSp-714626.
\end{acknowledgments}

\end{document}